\documentstyle[12pt,epsfig,psfig]{article}
\begin{document}

\def\gsim{{~\raise.15em\hbox{$>$}\kern-.85em
          \lower.35em\hbox{$\sim$}~}}
\def\lsim{{~\raise.15em\hbox{$<$}\kern-.85em
          \lower.35em\hbox{$\sim$}~}}

\begin{titlepage}
\title{\Large{Constraining Extensions of the Quark Sector\\
with the CP Asymmetry in  $B\rightarrow\psi K_{S}$}}
\author{Galit Eyal\thanks{e-mail: galit@wicc.weizmann.ac.il}\,\, and 
 Yosef Nir\thanks{e-mail: ftnir@wicc.weizmann.ac.il}\\
\small{Department of Particle Physics, Weizmann Institute of Science,
Rehovot 76100, Israel}}
\date{\small{WIS-99/28/AUG-DPP}}
\maketitle

\noindent
Models with extended quark sector affect the CP asymmetry in the 
$B\rightarrow \psi K_{S}$ decay, $a_{\psi K_{S}}$, in two ways: 
First, the top-mediated box diagram is not necessarily the only important 
contribution to $B-\bar{B}$ mixing. Second, the $3\times 3$ CKM matrix is 
no longer unitary. We analyze the constraints that follow from the CDF 
measurement, $a_{\psi K_{S}}=0.79^{+0.41}_{-0.44}$, on the mixing
parameters of extended quark sectors. Most noticeably, we find
significant constraints on the phase of the relevant flavor changing 
$Z$ coupling in models with extra down quarks in vector-like
representations. Further implications for the CP asymmetry in semileptonic 
$B$ decays are discussed.

\end{titlepage}

\section{Introduction}
The CDF collaboration has reported a preliminary measurement of
the CP asymmetry in the $B\rightarrow\psi K_{S}$ decay~\cite{a:cdf99}:
\begin{equation}\label{CDFval}
a_{\psi K_S}=0.79^{+0.41}_{-0.44},
\end{equation}
where
\begin{equation}  
\frac{\Gamma(\bar{B}^{0}_{\rm phys}(t)\rightarrow\psi K_{S})-
\Gamma(B^{0}_{\rm phys}(t)\rightarrow\psi K_{S})}
{\Gamma(\bar{B}^{0}_{\rm phys}(t)\rightarrow\psi K_{S})+
\Gamma(B^{0}_{\rm phys}(t)\rightarrow\psi K_{S})}=a_{\psi
K_{S}}\sin(\Delta m_{B}t).
\end{equation}
(Previous searches have been reported by OPAL~\cite{a:opal98} and by
CDF~\cite{a:cdf98}.) Within the Standard Model (SM), this value is cleanly
interpreted in terms of the angle $\beta$ of the unitarity triangle,
$a_{\psi K_S}=\sin 2\beta$. In the presence of new physics, this
interpretation is modified.

In a previous work with Barenboim~\cite{a:ben99}, we analyzed the
constraints that follow from eq.~(\ref{CDFval}) on the size and,
in particular, the phase of contributions from new physics to $B-\bar{B}$ 
mixing. There we investigated models in which the only relevant effect is 
a new, significant contribution to $B-\bar{B}$ mixing.
This large class of models includes, for example, supersymmetric and
left-right symmetric extensions of the SM. In particular, we assumed that the 
$b\rightarrow c\bar{c}s$ decay is dominated by the $W$-mediated tree-level 
diagram and that the $3\times3$ CKM matrix is unitary. While the first 
ingredient holds in almost all reasonable extensions of the SM, the second 
holds only in models where the quark sector consists of just the three 
generations of the SM.  In this work, we study extensions of the quark sector, 
namely we relax the assumption that the $3\times 3$ CKM matrix is unitary.
 
While in ref.~\cite{a:ben99} the analysis was (within the stated assumptions)
model-independent, here only the formalism is the same for all models. 
We introduce this formalism in section 2. To get numerical results
we have to separately discuss sequential and non-sequential extra down quarks.
We discuss models with extra down quarks in vector-like representations in 
section 3 and analyze the four generation model in section 4.
We summarize our results in section 5.

\section{Violation of CKM Unitarity and $B-\bar B$ Mixing}
We consider models where the new physics
does not contribute significantly to $W$-mediated tree level processes.
Most well-motivated extensions of the SM belong to this class.
The SM-dominance of these decays has three relevant consequences:
\begin{itemize}
\item[(i)] The phase of the $b\rightarrow c\bar cs$ 
decay amplitude, $A_{c\bar cs}$, is the CKM phase, $\arg(V_{cb}V_{cs}^*)$. 
\item[(ii)] The absorptive part of the $B-\bar{B}$ mixing amplitude
is not significantly modified by the new physics, $\Gamma_{12}\approx
\Gamma_{12}^{\rm SM}$.
\item[(iii)] The following measurements of CKM parameters are valid in 
our framework~\cite{a:pdg98}:
 \begin{eqnarray}\label{CKMcon}
 |V_{ud}|=0.9740\pm 0.0010,& \hspace{1cm}&|V_{cd}|=0.224\pm 0.016,\nonumber\\
 |V_{us}|=0.2196\pm 0.0023,& &|V_{cs}|=1.04\pm 0.16,\\ 
 \left|V_{ub}/V_{cb}\right|=0.08\pm 0.02,& &|V_{cb}|=0.0395\pm 0.0017.\nonumber
 \end{eqnarray}
\end{itemize}
The ranges in eq.~(\ref{CKMcon}) lead to the following bound, which 
plays a role in our discussion below:  
\begin{equation}\label{Rubound}
 R_{u}\equiv\left|\frac{V_{ud}V_{ub}^*}{V_{cd}V_{cb}^*}\right|\lsim0.47.
\end{equation}

We note that for processes where the SM tree and penguin contributions are
comparable, such as $b\rightarrow u\bar us$ decays, the new physics
contributions could be significant. This is the reason that we do not use
the bounds on the angle $\gamma$ of the unitarity triangle that follow from
$B\rightarrow \pi K$ decays~\cite{a:n99}. 

\begin{figure}[htb]
\begin{center}
\mbox{\epsfig{figure=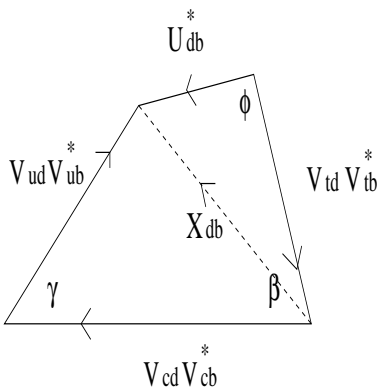,angle=0,width=5cm,height=5cm}}
\end{center}
\caption{The unitarity quadrangle.}
\label{f:nouni}
\end{figure}

Our investigation concerns models where the $3\times3$ CKM matrix is not 
unitary. In particular, we are interested in violation of the relation
$V^{*}_{ud}V_{ub}+V^{*}_{cd}V_{cb}+V^{*}_{td}V_{tb}=0$.
In any given model of this class, we can define a quantity $U_{db}$ such that
\begin{equation}\label{defUdb}
 V^{*}_{ud}V_{ub}+V^{*}_{cd}V_{cb}+V^{*}_{td}V_{tb}-U_{db}=0.
 \end{equation}
The physical interpretation of $U_{db}$ will be model-dependent.
However, in all models it gives a useful parameterization of
both the new physics contribution to $B-\bar{B}$ mixing and the
violation of CKM unitarity. In particular, it is convenient to
discuss the violation of CKM unitarity in terms of the
{\it unitarity quadrangle} drawn in fig.~\ref{f:nouni}.
 
We find it convenient to define also the following quantity
(see fig.~\ref{f:nouni}\ for its geometrical interpretation):
\begin{equation}\label{defXdb}
 X_{db}=U_{db}^{*}-V_{td}V^{*}_{tb}.
\end{equation}
We can bound $|X_{db}|$ through
\begin{equation}
 |V_{cd}V_{cb}^{*}|_{\rm min}-|V_{ud}V_{ub}^{*}|_{\rm max} \leq |X_{db}| \leq 
 |V_{cd}V_{cb}^{*}|_{\rm max}+|V_{ud}V_{ub}^{*}|_{\rm max}. \label{eq:xlim}
 \end{equation}
Using eq.~(\ref{CKMcon}) we get
\begin{equation}\label{Xdbbound}
0.004\lsim|X_{db}|\lsim0.014.
\end{equation}

The CP asymmetry in $B\rightarrow\psi K_S$ depends on the relative phase
between the $B-\bar{B}$ mixing amplitude and the $b\rightarrow c\bar{c}s$
decay amplitude. In our framework, neither $A_{c\bar cs}$ nor
$\Gamma_{12}$ are significantly affected by the new physics.
However, the new physics may give significant contributions to the 
dispersive part of the mixing amplitude, $M_{12}$.
The modification of $M_{12}$ can be parameterized as follows (see for
example~\cite{{a:babar},{a:gnw97}}): 
\begin{equation}
M_{12}=r_{d}^{2} e^{2i\theta_{d}}M_{12}^{\rm SM}.
\end{equation}
At present we have two experimental probes of $M_{12}$. 
The mass difference between the two neutral $B$ mesons, $\Delta m_B$,
is given by 
\begin{equation}
\Delta m_B=2r_d^2|M_{12}^{\rm SM}|,
\end{equation} 
so that its experimental value gives:
 \begin{equation}\label{eq:limrd}
 7.1\times 10^{-3} \leq r_{d}|V_{td}V_{tb}^{*}|\leq 10.75\times 10^{-3}.
 \end{equation}
The CP asymmetry in $B\rightarrow\psi K_S$, $a_{\psi K_S}$, is given by
\begin{equation}
a_{\psi K_{S}}=\sin [2(\beta+2\theta_{d})],
\end{equation}
so that its experimental value gives:
 \begin{equation}
 \sin [2(\beta+\theta_{d})]\gsim\cases{0.35&
one sigma,\cr 0&95\% CL.}\label{eq:limthe1}
 \end{equation} 

The CDF measurement constrains the combination $\beta+\theta_d$ 
through eq.~(\ref{eq:limthe1}). We are interested in finding whether the
range of the phase $\theta_d$ is constrained. We find that this is
indeed the case in models where the following situation holds:
First, there should be an independent upper bound on $\beta$,  
$|\beta|\leq \beta_{\rm max}$. Second, this upper bound should be strong 
enough in the sense that the following inequality holds:
\begin{equation}
\beta_{\rm max} < \frac{1}{2}\left(\arcsin (a_{\psi K_{S}})+\frac{\pi}{2}
\right).
\label{eq:bmaxc}
\end{equation}
Then, the following limit on $\sin 2\theta_{d}$ holds:
\begin{equation}
\sin(2\theta_{d})>\sin(\arcsin (a_{\psi K_{S}})-2\beta_{\rm max}).
\label{eq:stdcon}
\end{equation}

If CKM unitarity is not violated, then $\beta_{\rm max}\approx\pi/6$,
leading to $\sin 2\theta_{d}\gsim-0.6(-0.87)$ at one sigma (95\%
CL)~\cite{a:ben99}. If, however, CKM unitarity is violated, then
there is no model-independent constraint on $|V_{td}V_{tb}^{*}|$ and/or 
$\beta$ and we cannot constrain $r_{d}$ or $\theta_{d}$ without further input. 
We will constrain $r_{d}$ and $\theta_{d}$ in the framework of specific models
in the next two sections.
 
The $r_{d}$ and $\theta_{d}$ parameters are related also to other physical 
observables. For instance, the CP asymmetry in semileptonic decays, 
$a_{\rm SL}$, is given by:
\begin{equation}
a_{\rm SL}={\cal I}m\frac{\Gamma_{12}}{M_{12}}.
\end{equation}
Since $(\Gamma_{12}/M_{12})^{\rm SM}$ is real to a good approximation, the
effects of new physics within our framework can be written as follows (for
more details see ref.~\cite{a:ben99}):
\begin{equation}
\frac{a_{\rm SL}}{(\Gamma_{12}/M_{12})^{\rm SM}}=-\frac{\sin
2\theta_{d}}{r^{2}_{d}}. \label{eq:asl}
\end{equation}
Our analysis allows us to constrain $a_{\rm SL}$ within these specific models.

As mentioned above, the special point about extensions of the quark sector
is that the violation of CKM unitarity and the new contributions to
$B-\bar B$ mixing are related. Specifically, if we define
\begin{equation}\label{eq:rdef}
re^{i\phi}\equiv\frac{U_{db}^{*}}{V^{*}_{tb}V_{td}},
\end{equation}
then we have, in general,
\begin{equation}\label{eq:m12ed}
r_{d}^{2}e^{2i\theta_{d}}=1+are^{-i\phi}-br^{2}e^{-2i\phi}.
\end{equation} 
The $a$ and $b$ parameters are model-dependent.

To understand the consequences of eqs.~(\ref{eq:rdef}) and
(\ref{eq:m12ed}), note that eq.~(\ref{Xdbbound}) gives bounds on
\begin{equation}\label{Xdbrp}
|X_{db}|=(1+r^2-2r\cos\phi)^{1/2}|V_{td}V_{tb}^*|,
\end{equation}
while eq.~(\ref{eq:limrd}) gives bounds on
\begin{equation}
r_{d}|V_{td}V_{tb}^{*}|=(1+2ar\cos\phi-2br^{2}\cos2\phi-2abr^{3}\cos\phi
+a^{2}r^{2}+b^{2}r^{4})^{1/4}|V_{td}V_{tb}^{*}|.
\end{equation}
The fact that the two constraints have to be satisfied for the same
value of $|V_{td}V_{tb}^{*}|$ may exclude regions in the $(\phi,r)$ plane.

\section{Extra SU(2)-Singlet Down Quarks} 
We consider a model with extra down quarks in a vector-like 
representation of the SM gauge group, $G_{SM}=SU(3)_{C}\times
SU(2)_{L}\times U(1)_{Y}$. In addition to the three quark generations,
each consisting of the three representations
\begin{equation}\label{eq:partic}
Q_{Li}(3,2)_{+1/6},\;\;\; u_{Ri}(3,1)_{+2/3},\;\;\; d_{Ri}(3,1)_{-1/3},
\;\;\; (i=1,2,3),
\end{equation}
we have the following vector-like representation:
\begin{equation}
d_{4}(3,1)_{-1/3}\; + \; \bar{d}_{4}(\bar{3},1)_{+1/3}.
\end{equation}
Such quark representations appear, for example, in $E_{6}$ GUTs.

The most important feature of this model to our purposes is that 
it allows for flavor changing $Z\bar{d}b$-couplings  (for details 
see refs.~\cite{a:ns901}-\cite{a:bbp95}):
\begin{equation}\label{LZdb}
{\cal L}_{Zdb}=-\frac{g}{2\cos\theta_W}U_{db}Z_\mu\bar d_L\gamma^\mu b_L+
{\rm h.c.}.
\end{equation}
The $U_{db}$ mixing parameter in eq.~(\ref{LZdb}) is the same parameter
defined in eq.~(\ref{defUdb}) which signifies violation of CKM unitarity.
It allows $Z$-mediated tree level contributions to flavor changing
neutral current processes such as $B\rightarrow\mu^+\mu^-X_d$.
The experimental bound on the rate of this decay gives 
(see {\it e.g.}~\cite{a:gnr98}\ and references therein):
\begin{equation}
\left|{U_{db}/V_{cb}}\right|\leq 0.04 \hspace{0.5cm} \rightarrow 
\hspace{0.5cm} |U_{db}|\leq 0.0016.\label{eq:ubdlim}
\end{equation}

Putting the bound on $|U_{db}|$ of eq.~(\ref{eq:ubdlim}) and the values
of the CKM parameters of eq.~(\ref{CKMcon}) into
\begin{equation}\label{eq:vtdvtb}
|V_{cd}V_{cb}^{*}|_{\rm min}-|V_{ud}V_{ub}^{*}|_{\rm max}-|U_{db}|_{\rm max}
\leq |V_{td}V^{*}_{tb}| \leq |V_{cd}V_{cb}^{*}|_{\rm max}+
|V_{ud}V_{ub}^{*}|_{\rm max}+|U_{db}|_{\rm max},
\end{equation}
we get
\begin{equation}\label{vtdvtb}
2.7\times10^{-3}\leq|V_{td}V^{*}_{tb}|\leq1.6\times10^{-2}.
\end{equation}
Using eqs.~(\ref{vtdvtb}) and (\ref{eq:limrd}), we can constrain $r_d$: 
\begin{equation}\label{eq:mdrd}
0.44 \lsim r_{d} \lsim 4.1. 
\end{equation}

Putting eqs.~(\ref{eq:ubdlim}) and (\ref{CKMcon}) into
\begin{equation}\label{sinbmax}
|\sin\beta|\leq\frac{|V_{ud}V_{ub}^{*}|_{\rm max}+|U_{db}|_{\rm max}}
{|V_{cd}V_{cb}^{*}|_{\rm min}}=(R_u)_{\rm max}+\frac{1}{|V_{cd}|_{\rm min}}
\left|\frac{U_{db}}{V_{cb}^{*}}\right|_{\rm max},
\end{equation}
we get
\begin{equation}\label{eq:sbmax}
\beta_{\rm max} \approx \frac{2\pi}{9}.
\end{equation}
Using eqs.~(\ref{eq:sbmax}) and (\ref{eq:limthe1}), we can constrain
$\theta_d$:
\begin{equation}\label{eq:mdtd}
\sin 2\theta_{d} \gsim\cases{-0.88&one sigma,\cr -0.99&95\% CL.\cr}
\end{equation}

In the derivation of (\ref{eq:mdrd}) and (\ref{eq:mdtd}), 
we have not used the correlation between violation of CKM unitarity
and contribution to $B-\bar B$ mixing. To do so, we note
that the $U_{db}$ coupling of eq.~(\ref{LZdb}) allows a $Z$-mediated 
tree diagram contribution to $B-\bar{B}$ mixing. It is possible to 
parameterize the new contributions to $M_{12}$ as in
eq.~(\ref{eq:m12ed})~\cite{{a:bbp95},{a:bb97},{a:bbbv97}}, with
\begin{equation}
a=\frac{4\bar{C}(x_{t})}{\bar{E}(x_{t})},\ \ \ 
b=\frac{4\pi\sin^{2}\theta_{W}}{\alpha\bar{E}(x_{t})}.
\end{equation}
Here $x_{t}=(m_{t}/M_{W})^{2}$ and $\bar{C}(x_t)$ and $\bar{E}(x_t)$
are the Inami-Lim functions~\cite{a:il81}:
\begin{eqnarray}
\bar{E}(x_{t})&=&\frac{-4x_{t}+11x_{t}^{2}-x_{t}^{3}}{4(1-x_{t})^{2}}
+\frac{3x_{t}^{3}\ln x_{t}}{2(1-x_{t})^{3}}, 
\label{eq:Eil}\\
\bar{C}(x_{t})&=&\frac{x_{t}}{4}\left[\frac{4-x_{t}}{1-x_{t}}+
\frac{3x_{t}\ln x_{t}}{(1-x_{t})^{2}}\right].
\end{eqnarray}
An explicit calculation~\cite{a:bb97}\ gives $a=-3.3$ and $b=-160$. 

\begin{figure}[htb]
\centerline{$(a)$}
\begin{center}
\mbox{\psfig{figure=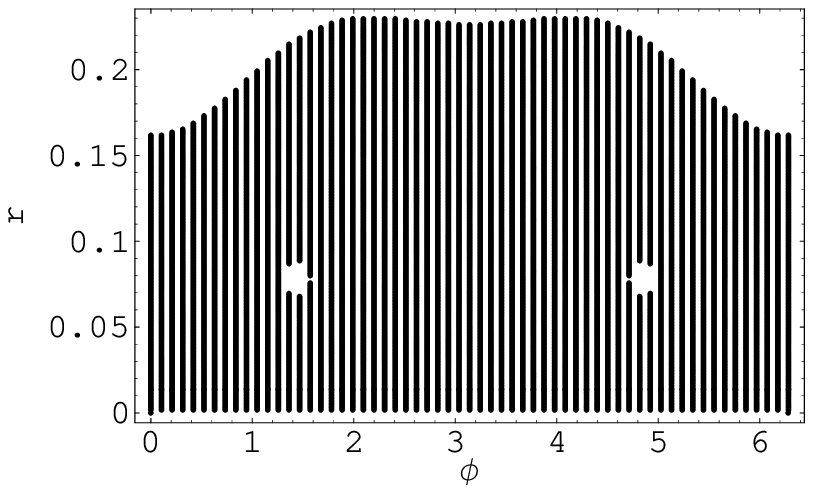,angle=0,width=8cm,height=4cm}}
\end{center}
\centerline{$(b)$}
\begin{center}
\mbox{\psfig{figure=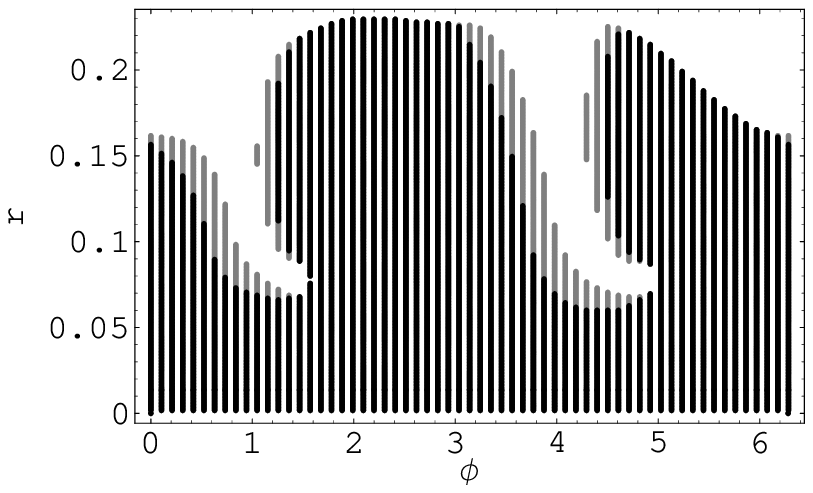,angle=0,width=8cm,height=4cm}}
\end{center}
\caption{SU(2)-singlet down quarks. ($a$) The $\Delta m_{B}$ constraint. 
The dark region is allowed. ($b$) The combination of the $\Delta m_{B}$ and 
$a_{\psi K_{S}}$ constraints. The dark (light plus dark) grey region is the 
allowed region corresponding to the one sigma ($95\%$ CL) bound, 
$a_{\psi K_{S}}\geq 0.35\; (0)$.} 
\label{f:rphiedm} 
\end{figure}

We now relate all of our constraints to the $(\phi,r)$ parameters by
performing the following procedure. We scan the $\phi-r$ parameter space. 
For each $(\phi,r)$ pair we calculate $r_{d}$ and $\theta_d$ 
through eq.~(\ref{eq:m12ed}) and $|X_{db}/(V_{td}V_{tb}^{*})|$ through 
eq.~(\ref{Xdbrp}). We check whether there exist values of $|V_{td}V_{tb}^{*}|$ 
that are consistent with all the constraints in eqs.~(\ref{Xdbbound}), 
(\ref{eq:limrd}), (\ref{eq:rdef}) and (\ref{eq:ubdlim}). Note that at this 
stage we do not yet incorporate the $a_{\psi K_S}$ constraint. The allowed 
region is displayed in fig.~\ref{f:rphiedm}($a$).
The upper bound on $r$ is a result of inconsistency between the
upper bound of eq.~(\ref{eq:limrd}) and the lower bound of
eq.~(\ref{Xdbbound}). The small excluded regions at $\cos\phi\approx0$
correspond to $r\approx|b|^{-1/2}$, where $r_d$ is too small. 

Next we incorporate the constraint from $a_{\psi K_S}$. For each pair
of values $(\phi,r)$, we calculate $\beta_{\rm max}$:
\begin{equation}
\beta_{\rm max}=\left| \arcsin\left(
\frac{r\sin\phi}{(1-2r\cos\phi+r^2)^{1/2}} \right)\right|+
\left[\arccos\left(\frac{|V_{cd}V_{cb}^{*}|^{2}+|X_{db}|^{2}- 
|V_{ud}V_{ub}^{*}|^{2}}{2|V_{cd}V_{cb}^{*}||X_{db}|}\right)\right]_{\rm max}.
\end{equation}
If the condition in eq.~(\ref{eq:bmaxc}) holds (and $r\cos\phi <1$), 
we exclude $(\phi,r)$ pairs that violate the bound in eq.~(\ref{eq:stdcon}). 
The allowed parameter space is displayed in fig.~\ref{f:rphiedm}($b$).
 
Our numerical scan gives, at the one sigma level:
\begin{eqnarray}
-0.8 \lsim &\sin2\theta_{d}& \leq 1,\\
0.5 \lsim &r_{d}& \lsim 3.2.
\end{eqnarray}\\
For the semileptonic asymmetry we find (see eq.~(\ref{eq:asl})):
\begin{equation}
-4.0 \lsim \frac{a_{SL}}{(\Gamma_{12}/M_{12})^{SM}} \lsim 1.4.
\end{equation}

\section{A Fourth Generation}
The second model we consider is the SM with a fourth sequential
generation. Here the quark content is as given in 
eq.~(\ref{eq:partic}) with $i=1,2,3,4$. 
A fourth generation by itself is now excluded by electroweak precision
data~\cite{a:pdg98}. However, if there is new physics in addition to a
fourth generation, such that the electroweak precision data constraints
are relaxed but $M_{12}$ is not affected by this extra new physics, then
our analysis below applies. The analysis in this section applies also to
a model in which extra up quarks in a vector-like representation,
\begin{equation}
u_{4}(3,1)_{+2/3}\; + \; \bar{u}_{4}(\bar{3},1)_{-2/3},
\end{equation} 
are added to the SM three generations~\cite{a:bbp95}.

Within these models,
\begin{equation}
U_{db}=-V^{*}_{t^\prime d}V_{t^\prime b}.
\end{equation}
From unitarity of the $4\times 4$ matrix, we have~\cite{{a:bc86},{a:hl86}}:
\begin{equation}
|V^{*}_{td}V_{tb}|\leq 0.1, \hspace{1cm} 
|U_{db}|\leq 0.1. \label{eq:ubdun}
\end{equation}
The reasonable agreement of the most recent data on
$R_b=\Gamma(Z\rightarrow b\bar b)/\Gamma(Z\rightarrow{\rm hadrons})$
and the expectation of the three generation SM implies that~\cite{a:babar}
\begin{equation}\label{Zbbcon}
1\simeq |V_{tb}|^{2}+|V_{t^\prime b}|^{2}
\left(\frac{m_{t^\prime}}{m_{t}}\right)^{2}.
\end{equation}
For $m_{t^\prime}=500\;GeV$,  eq.~(\ref{Zbbcon}) leads to 
\begin{equation}\label{ZbbUdb}
|U_{db}|\lsim 0.03.
\end{equation}
In our analysis, we use the bound (\ref{ZbbUdb}).

The analysis goes along similar lines to that of the previous section.
However, the fact that eq.~(\ref{ZbbUdb}) gives
$|U_{db}|_{\rm max}>|V_{cd}V_{cb}^{*}|_{\rm min}$ means that we can put 
neither a meaningful lower bound on $|V_{td}V^{*}_{tb}|$ nor a meaningful 
upper bound on $|\sin\beta|$. Instead of (\ref{vtdvtb}), we now have
\begin{equation}\label{vtdvtbf}
|V_{td}V^{*}_{tb}|\leq0.044,
\end{equation}
and, consequently, eq.~(\ref{eq:limrd}) gives only a lower bound on $r_d$: 
\begin{equation}\label{eq:mdrdf}
r_{d} \gsim0.16. 
\end{equation}
As concerns $\sin2\theta_d$, we get no bounds. 

Next, we want to incorporate the relation between violation of CKM unitarity
and new contributions to $B-\bar B$ mixing. In the four generation model, 
the new contributions to $B-\bar{B}$ mixing come from box diagrams involving 
$t^\prime$ quarks. These contributions can again be parameterized as in 
eq.~(\ref{eq:m12ed})~\cite{{a:whhh91},{a:hhw98}} with
\begin{equation}
a=-2\frac{\bar{E}(x_{t^\prime},x_{t})}{\bar{E}(x_{t})},\ \ \ 
b=-\frac{\bar{E}(x_{t^\prime})}{\bar{E}(x_{t})},
\end{equation}
\begin{eqnarray}
\bar{E}(x_{t^\prime},x_{t})&=&-x_{t^\prime}x_{t}\left[
\frac{1}{x_{t^\prime}-x_{t}}\left(\frac{1}{4}-\frac{3}{2}
\frac{1}{(x_{t^\prime}-1)}-\frac{3}{4}\frac{1}{(x_{t^\prime}-1)^{2}}
\right)\ln x_{t^\prime}\right.\nonumber\\
& &+\left.(x_{t^\prime}\leftrightarrow x_{t})-\frac{3}{4}\frac{1}
{(x_{t^\prime}-1)(x_{t}-1)}\right], 
\end{eqnarray}
and $\bar{E}(x)$ defined in eq.~(\ref{eq:Eil}).
Taking $m_{t}\approx170\;GeV$ and $180\;GeV \lsim m_{t^\prime}\lsim500\;GeV$, 
we find: 
\begin{equation}
-4\lsim a \lsim -2, \hspace{2cm} -5.5 \lsim b \lsim -1.
\end{equation}

\begin{figure}[htb]
\centerline{$(a)$}
\begin{center}
\mbox{\psfig{figure=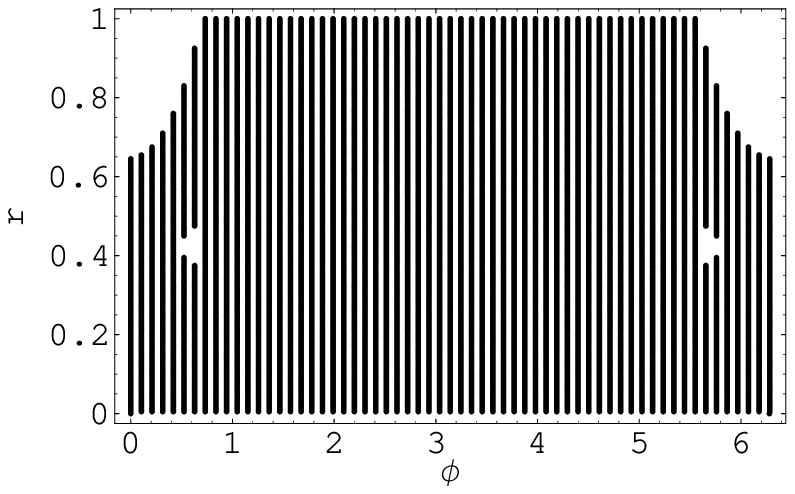,angle=0,width=8cm,height=4cm}}
\end{center}
\centerline{$(b)$}
\begin{center}
\mbox{\psfig{figure=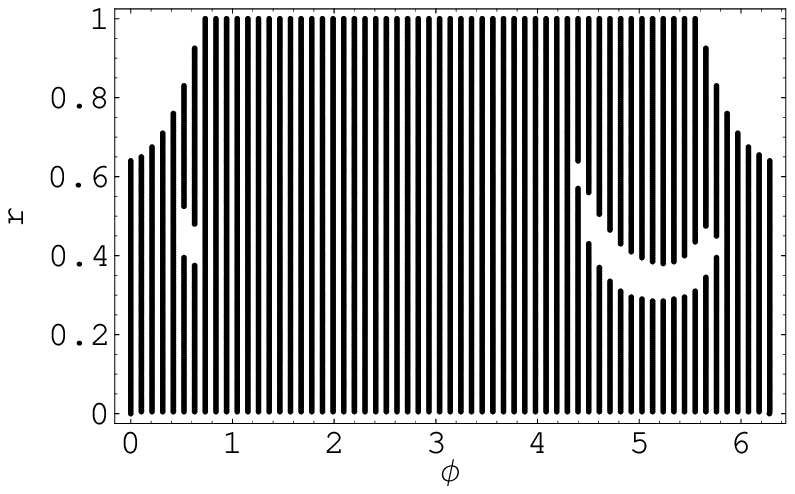,angle=0,width=8cm,height=4cm}}
\end{center}
\caption{Four generations, $m_{t^\prime}=500\;GeV$. The dark region is allowed.
($a$) The $\Delta m_{B}$ constraint. ($b$) The combination of the
$\Delta m_{B}$ and $a_{\psi K_{s}}$ one sigma constraints.}
\label{f:foca21}
\end{figure}

Below we display only the results of a numerical analysis for the case
$m_{t^\prime}=500\;GeV$ for which the effects are most significant. In this
case: $a\sim-3.8$ and $b\sim-5.4$. The allowed region without the 
$a_{\psi K_S}$ constraint is given in fig.~\ref{f:foca21}($a$). We only display
the $r<1$ region since the $a_{\psi K_S}$ constraint will have no effect
for $r>1$. The excluded region around $\phi=0$ is a result of inconsistency 
between the upper bound of eq.~(\ref{eq:limrd}) and the lower bound of 
eq.~(\ref{Xdbbound}). The small excluded regions at $r\approx0.4$
correspond to $r_d$ values that are too small. Note that $r_d$ can be
very large in this model, corresponding to a nearly vanishing
$|V_{td}V_{tb}^*|$ and, consequently, nearly vanishing $M_{12}^{\rm SM}$. 
Incorporating the $a_{\psi K_S}$ constraint, we find that the new CDF 
measurement does not place significant new constraints on the parameter 
space of the four generation model. This is particularly true for a 
$t^\prime$-mass that is not much higher than $m_t$.  The allowed region
is displayed in fig.~\ref{f:foca21}($b$).  

Our numerical scan gives
\begin{equation}
r_{d}\gsim 0.33,
\end{equation}
and no bounds on $\sin2\theta_d$. For the semileptonic asymmetry, we get
\begin{equation}
-9.0 \lsim \frac{a_{\rm SL}}{(\Gamma_{12}/M_{12})^{\rm SM}} \lsim 6.1.
\end{equation}

\section{Conclusions}
In a previous work with Barenboim~\cite{a:ben99}, we used the new
CDF measurement of the CP asymmetry in $B\rightarrow\psi K_S$ to derive 
the first constraint on the phase of new physics contributions to
$B-\bar B$ mixing. We have done so in the framework of models where the
CKM matrix is unitary.
In this work we have shown that significant constraints apply also in
extensions of the quark sector, where the $3\times3$ CKM matrix is not
unitary. The main reason that makes this possible is that a single complex
parameter ($U_{db}$) characterizes both the violation of CKM unitarity and
the new contributions to $B-\bar B$ mixing. Therefore, the number of
relevant new parameters is effectively the same as in models where the CKM
matrix is unitary. In either case, the measurement of $a_{\psi K_S}$
gives the first constraint on the phase $2\theta_{d}=
\arg(M_{12}/M_{12}^{\rm SM})$. Specifically, whenever we can put an
upper bound on $|\beta|$ that is lower than $\frac{1}{2}
\left(\arcsin (a_{\psi K_{S}})+\frac{\pi}{2} \right)$,  it follows
that there is a lower bound on $\sin2\theta_d$.  

Our most significant results concern models with extra SU(2)-singlet
down quarks. The constraints on the relevant mixing parameters are
displayed in fig.~\ref{f:rphiedm}($b$). In particular, the measurement of
$a_{\psi K_S}$ gives $\sin 2\theta_{d}\gsim-0.8$. This phase
is related to the phase of $U_{db}$ which, in this framework, parametrizes 
the flavor changing $Z\bar db$ coupling. The bound on $\sin2\theta_d$
together with constraints from $\Delta m_{B}$ give bounds on the CP asymmetry 
in semileptonic B decays, $-1.4\times 10^{-2} \lsim a_{\rm SL} \lsim 
4.0\times 10^{-2}$.
Weaker constraints apply to the four generation model and to models
with extra up quarks in vector-like representations.

\bigskip
\noindent
{\bf Acknowledgements:}
\smallskip
\noindent
We thank JoAnne Hewett for useful discussions.
Y.N. is supported in part by the United States $-$ Israel Binational
Science Foundation (BSF), by the Israel Science Foundation founded
by the Israel Academy of Sciences and Humanities
and by the Minerva Foundation (Munich).

{}
\end{document}